\documentclass[preprint,prd,aps,showpacs,showkeys,nofootinbib]{revtex4}
\usepackage{graphicx}
\begin{document}

%\begin{CJK*}{GB}{gbsn}

%\fancyhead[c]{\small Chinese Physics C~~~Vol. xx, No. x (201x) xxxxxx}
%\fancyfoot[C]{\small 010201-\thepage}

\title{125 GeV Higgs decay with lepton flavor violation in the $\mu\nu$SSM}

\author{Hai-Bin Zhang$^{1,}$\footnote{hbzhang@hbu.edu.cn},
Tai-Fu Feng$^{1,2,}$\footnote{fengtf@hbu.edu.cn},
Shu-Min Zhao$^{1}$,\\
Yu-Li Yan$^{1}$,
Fei Sun$^{3}$}

\affiliation{$^1$ Department of Physics, Hebei University, Baoding, 071002, China\\
$^2$ State Key Laboratory of Theoretical Physics (KLTP),\\
Institute of Theoretical Physics, Chinese Academy of Sciences, Beijing, 100190, China\\
$^3$ College of Science, China Three Gorges University, Yichang, 443002, China}

\begin{abstract}
Recently, the CMS and ATLAS Collaborations have reported direct searches for the 125 GeV Higgs decay with lepton flavor violation, $h\rightarrow \mu \tau$. In this work, we analyze the signal of the lepton flavour violating (LFV) Higgs decay $h\rightarrow \mu \tau$ in the $\mu$ from $\nu$ Supersymmetric Standard Model ($\mu\nu$SSM) with slepton flavor mixing. Simultaneously, we consider the constraints from the LFV decay $\tau \rightarrow \mu \gamma$, the muon anomalous magnetic dipole moment and the lightest Higgs mass around 125 GeV.
\end{abstract}


\maketitle

\section{Introduction\label{sec1}}
The discovery of the Higgs boson by the ATLAS and CMS Collaborations~\cite{ATLAS,CMS} is a great success of the Large Hadron Collider~(LHC). Combining the updated data of the ATLAS and CMS Collaborations, the measured mass of the Higgs boson now is~\cite{ATLAS-CMS}
\begin{eqnarray}
m_h=125.09\pm 0.24\: {\rm{GeV}}.
\end{eqnarray}
The next step is focusing on searching for its properties. In the Standard Model (SM), which is renormalizable, lepton flavour violating (LFV) Higgs decays are forbidden~\cite{Harnik}. But recently, a direct search for the 125 GeV Higgs decay with lepton flavor violation, $h\rightarrow \mu \tau$, has been described by the CMS Collaboration~\cite{CMS-LFVH,CMS-LFVH-1}. The upper limit on the branching ratio of $h\rightarrow \mu \tau$ at 95\% confidence level (CL) is~\cite{CMS-LFVH-1}
\begin{eqnarray}
{\rm{Br}}(h\rightarrow \mu \tau)<1.20\times 10^{-2}.
\label{uplimit}
\end{eqnarray}
Here, interpreted as a signal, $\mu \tau$ means the final state consisting of $\bar{\mu} \tau$ and $\mu \bar{\tau}$.

The ATLAS Collaboration gives the constraint on the branching ratio of $h\rightarrow \mu \tau$ at 95\% CL to be~\cite{ATLAS-LFVH,ATLAS-LFVH-1}
\begin{eqnarray}
{\rm{Br}}(h\rightarrow \mu \tau)<1.43\times 10^{-2}.
\end{eqnarray}
The ATLAS and CMS experiments do not currently show a significant deviation from the SM. Therefore, the experiments still need to make more precise measurements in the future.

LFV Higgs decays can occur naturally in models beyond the SM, such as supersymmetric models~\cite{LFVHD-Br1,LFVHD2,LFVHD2-0,LFVHD2-1,LFVHD2-2,LFVHD-Br2,LFVHD2-3,LFVHD2-4,LFVHD3}, composite Higgs boson models~\cite{LFVHD4,LFVHD5}, Randall-Sundrum models~\cite{LFVHD6,LFVHD7,LFVHD8}, and many others ~\cite{LFVHD9,LFVHD10,LFVHD11,LFVHD12,LFVHD13,LFVHD14,LFVHD15,LFVHD15-1,LFVHD45,
LFVHD46,LFVHD47,LFVHD48,LFVHD49,LFVHD50,LFVHD51,LFVHD60,LFVHD61,LFVHD62}.
Due to the running of the LHC, LFV Higgs decays have recently been discussed within various theoretical frameworks~\cite{LFVHD20,LFVHD21,LFVHD22,LFVHD23,LFVHD24,LFVHD25,LFVHD26,LFVHD27,LFVHD28,LFVHD28-1,LFVHD29,LFVHD30,LFVHD31,
LFVHD32,LFVHD33,LFVHD34,LFVHD35,LFVHD36,LFVHD37,LFVHD38,LFVHD39,LFVHD40,LFVHD41,LFVHD42,LFVHD42-1,LFVHD42-2,LFVHD43,LFVHD44,
LFVHD52,LFVHD53,LFVHD54,LFVHD55,LFVHD56,LFVHD57,LFVHD58,LFVHD59,
LFVHD63,LFVHD64,LFVHD65,LFVHD66,LFVHD67,LFVHD68,LFVHD69,LFVHD70,LFVHD71,LFVHD72,LFVHD73,LFVHD74,LFVHD75,LFVHD76,LFVHD77,
LFVHD78,LFVHD79,LFVHD80,LFVHD81,LFVHD82,LFVHD83,LFVHD84,LFVHD85,LFVHD86,LFVHD87,LFVHD88,LFVHD89,LFVHD90,LFVHD91}.
In this paper, we will study the LFV Higgs decay $h\rightarrow \mu \tau$  in the ``$\mu$ from $\nu$ Supersymmetric Standard Model'' ($\mu\nu$SSM)~\cite{mnSSM,mnSSM1,mnSSM2}. As an extension of the Minimal Supersymmetric Standard Model (MSSM)~\cite{MSSM,MSSM1,MSSM2,MSSM3,MSSM4}, the $\mu\nu$SSM solves the $\mu$ problem~\cite{m-problem} of the MSSM, through the R-parity breaking couplings ${\lambda _i}\hat \nu _i^c\hat H_d^a\hat H_u^b$ in the superpotential. The $\mu$ term is generated spontaneously via the nonzero vacuum expectative values (VEVs) of right-handed sneutrinos, $\mu  = {\lambda _i}\left\langle {\tilde \nu _i^c} \right\rangle$, when the electroweak symmetry is broken (EWSB). In addition, nonzero VEVs of sneutrinos in the $\mu\nu$SSM can generate three tiny massive Majorana neutrinos at tree level through the seesaw machanism~\cite{mnSSM,mnSSM1,mnSSM2,meu-m,meu-m1,meu-m2,meu-m3,neu-zhang1}.

Within the $\mu\nu$SSM, we have studied some LFV processes, $l_j^-\rightarrow l_i^-\gamma$, $l_j^-  \rightarrow l_i^- l_i^- l_i^+$,  muon conversion to electrons in nuclei and  $Z\rightarrow l_i^\pm l_j^\mp$ in our previous work~\cite{ref-zhang,ref-zhang1,ref-zhang-LFV}. The numerical results show that the LFV rates for $l_j-l_i$  transitions in the $\mu\nu$SSM depend on the slepton flavor mixing, and the present experimental limits for the branching ratio of $l_j^-\rightarrow l_i^-\gamma$ constrain the slepton mixing parameters most strictly~\cite{ref-zhang-LFV}. In this work, considering the constraint of $\tau \rightarrow \mu \gamma$, we continue to analyze the LFV Higgs decay $h\rightarrow \mu \tau$  in the $\mu\nu$SSM with slepton flavor mixing.

The paper is organized as follows. In Section~2, we briefly present the $\mu\nu$SSM, including its superpotential and the general soft SUSY-breaking terms. Section~3 contains the analytical expressions of the 125 GeV Higgs decay with lepton flavor violation in the $\mu\nu$SSM. The numerical analysis and the summary are given in Section~4 and Section~5, respectively. Some formulae are collected in Appendix~\ref{app-coupling} and Appendix~\ref{app-MDM}.

\section{The $\mu\nu$SSM\label{sec2}}
In addition to the superfields of the MSSM, the $\mu\nu$SSM introduces right-handed neutrino superfields $\hat{\nu}_i^c$~$(i=1,2,3)$. Besides the MSSM Yukawa couplings for quarks and charged leptons, the superpotential of the $\mu\nu$SSM contains Yukawa couplings for neutrinos, two additional types of terms involving the Higgs doublet superfields $\hat H_u$ and $\hat H_d$, and the right-handed neutrino superfields  $\hat{\nu}_i^c$,~\cite{mnSSM}
\begin{eqnarray}
&&\hspace{-1.0cm}W={\epsilon _{ab}}\left( {Y_{{u_{ij}}}}\hat H_u^b\hat Q_i^a\hat u_j^c + {Y_{{d_{ij}}}}\hat H_d^a\hat Q_i^b\hat d_j^c
+ {Y_{{e_{ij}}}}\hat H_d^a\hat L_i^b\hat e_j^c \right)  \nonumber\\
&&\hspace{-0.2cm}
+ {\epsilon _{ab}}{Y_{{\nu _{ij}}}}\hat H_u^b\hat L_i^a\hat \nu _j^c -  {\epsilon _{ab}}{\lambda _i}\hat \nu _i^c\hat H_d^a\hat H_u^b + \frac{1}{3}{\kappa _{ijk}}\hat \nu _i^c\hat \nu _j^c\hat \nu _k^c ,
\label{eq-W}
\end{eqnarray}
where $\hat H_u^T = \Big( {\hat H_u^ + ,\hat H_u^0} \Big)$, $\hat H_d^T = \Big( {\hat H_d^0,\hat H_d^ - } \Big)$, $\hat Q_i^T = \Big( {{{\hat u}_i},{{\hat d}_i}} \Big)$, $\hat L_i^T = \Big( {{{\hat \nu}_i},{{\hat e}_i}} \Big)$ are $SU(2)$ doublet superfields, and $\hat u_i^c$, $\hat d_i^c$, and $\hat e_i^c$ denote the singlet up-type quark, down-type quark and charged lepton superfields, respectively.  Here, $Y$, $\lambda$, and $\kappa$ are dimensionless matrices, a vector, and a totally symmetric tensor.  $i,j,k=1,2,3$ are the generation indices, $a,b=1,2$ are the SU(2) indices with antisymmetric tensor $\epsilon_{12}=1$. In the superpotential, the last two terms explicitly violate lepton number and R-parity. Note that the summation convention is implied on repeated indices in this paper.

Once EWSB occurs, the neutral scalars develop in general the VEVs:
\begin{eqnarray}
\langle H_d^0 \rangle = \upsilon_d , \quad \langle H_u^0 \rangle = \upsilon_u , \quad
\langle \tilde \nu_i \rangle = \upsilon_{\nu_i} , \quad \langle \tilde \nu_i^c \rangle = \upsilon_{\nu_i^c}.
\end{eqnarray}
Then, the terms ${\epsilon _{ab}}{Y_{{\nu _{ij}}}}\hat H_u^b\hat L_i^a\hat \nu _j^c$ and ${\epsilon _{ab}}{\lambda _i}\hat \nu _i^c\hat H_d^a\hat H_u^b$ in the superpotential can generate the effective bilinear terms $\epsilon _{ab} \varepsilon_i \hat H_u^b\hat L_i^a$ and $\epsilon _{ab} \mu \hat H_d^a\hat H_u^b$, with $\varepsilon_i= Y_{\nu _{ij}} \left\langle {\tilde \nu _j^c} \right\rangle$ and $\mu  = {\lambda _i}\left\langle {\tilde \nu _i^c} \right\rangle$.  One can define the neutral scalars as
\begin{eqnarray}
&&\hspace{-0.5cm}H_d^0=\frac{h_d + i P_d}{\sqrt{2}} + \upsilon_d, \quad\; \tilde \nu_i = \frac{(\tilde \nu_i)^\Re + i (\tilde \nu_i)^\Im}{\sqrt{2}} + \upsilon_{\nu_i},  \nonumber\\
&&\hspace{-0.5cm}H_u^0=\frac{h_u + i P_u}{\sqrt{2}} + \upsilon_u, \quad \tilde \nu_i^c = \frac{(\tilde \nu_i^c)^\Re + i (\tilde \nu_i^c)^\Im}{\sqrt{2}} + \upsilon_{\nu_i^c}.
\end{eqnarray}

In the framework of supergravity-mediated supersymmetry breaking, the general soft SUSY-breaking terms of the $\mu\nu$SSM are given by
\begin{eqnarray}
&&\hspace{-1.8cm}- \mathcal{L}_{soft}=m_{{{\tilde Q}_{ij}}}^{\rm{2}}\tilde Q{_i^{a\ast}}\tilde Q_j^a
+ m_{\tilde u_{ij}^c}^{\rm{2}}\tilde u{_i^{c\ast}}\tilde u_j^c + m_{\tilde d_{ij}^c}^2\tilde d{_i^{c\ast}}\tilde d_j^c
+ m_{{{\tilde L}_{ij}}}^2\tilde L_i^{a\ast}\tilde L_j^a  \nonumber\\
&&\hspace{0cm} +  m_{\tilde e_{ij}^c}^2\tilde e{_i^{c\ast}}\tilde e_j^c + m_{{H_d}}^{\rm{2}} H_d^{a\ast} H_d^a
+ m_{{H_u}}^2H{_u^{a\ast}}H_u^a + m_{\tilde \nu_{ij}^c}^2\tilde \nu{_i^{c\ast}}\tilde \nu_j^c \nonumber\\
&&\hspace{0cm}  +  \epsilon_{ab}\Big[{{({A_u}{Y_u})}_{ij}}H_u^b\tilde Q_i^a\tilde u_j^c
+ {{({A_d}{Y_d})}_{ij}}H_d^a\tilde Q_i^b\tilde d_j^c
\nonumber\\
&&\hspace{0cm} + {{({A_e}{Y_e})}_{ij}}H_d^a\tilde L_i^b\tilde e_j^c + {\rm{H.c.}} \Big] + \Big[ {\epsilon _{ab}}{{({A_\nu}{Y_\nu})}_{ij}}H_u^b\tilde L_i^a\tilde \nu_j^c
\nonumber\\
&&\hspace{0cm} - {\epsilon _{ab}}{{({A_\lambda }\lambda )}_i}\tilde \nu_i^c H_d^a H_u^b
+ \frac{1}{3}{{({A_\kappa }\kappa )}_{ijk}}\tilde \nu_i^c\tilde \nu_j^c\tilde \nu_k^c + {\rm{H.c.}} \Big] \nonumber\\
&&\hspace{0cm}  -  \frac{1}{2}\Big({M_3}{{\tilde \lambda }_3}{{\tilde \lambda }_3}
+ {M_2}{{\tilde \lambda }_2}{{\tilde \lambda }_2} + {M_1}{{\tilde \lambda }_1}{{\tilde \lambda }_1} + {\rm{H.c.}} \Big).
\end{eqnarray}
Here, the first two lines contain mass squared terms of squarks, sleptons, Higgses and sneutrinos. The next three lines include the trilinear scalar couplings. In the last line, $M_3$, $M_2$, and $M_1$ represent the Majorana masses corresponding to $SU(3)$, $SU(2)$, and $U(1)$ gauginos $\hat{\lambda}_3$, $\hat{\lambda}_2$, and $\hat{\lambda}_1$, respectively. In addition, the tree-level scalar potential receives the usual $D$- and $F$-term contributions~\cite{mnSSM1}.

In the $\mu\nu$SSM, the quadratic potential includes
\begin{eqnarray}
&&\hspace{-1.0cm}{V_{quadratic}} = \frac{1}{2} {S'^T}M_S^2S' + \frac{1}{2}{P'^T}M_P^2P' + {S'^{-T}}M_{S^\pm}^2{S'^+}
\nonumber\\
&&\hspace{0.8cm}+  (\frac{1}{2}{\chi '^{0 T}}{M_n}{\chi '^0 } +  {\Psi ^{ - T}}{M_c}{\Psi^+} + {\rm{H.c.}} )+ \cdots ,
\label{eq-quad}
\end{eqnarray}
where in the unrotated basis ${S'^T} = ({h_d},{h_u},{(\tilde \nu_i)^\Re},{({\tilde \nu_i^c})^\Re})$, ${P'^T} = ({P_d},{P_u},{(\tilde \nu_i)^\Im},{({\tilde \nu_i^c})^\Im})$, ${S'^{ \pm T}} = (H_d^ \pm ,H_u^ \pm ,\tilde e_{L_i}^ \pm ,\tilde e_{R_i}^ \pm )$, ${\Psi ^{ - T}} = \left( { - i{{\tilde \lambda }^ - },\tilde H_d^ - ,e_{L{_i}}^ - } \right)$, ${\Psi ^{ + T}} = \left( { - i{{\tilde \lambda }^ + },\tilde H_u^ + ,e_{R{_i}}^+} \right)$ and ${\chi '^{ 0 T}} = \left( {{{\tilde B}^0 },{{\tilde W}^0 },{{\tilde H}_d}{\rm{,}}{{\tilde H}_u},{\nu_{R_i}},{\nu_{L{_i}}}} \right)$.
The concrete expressions for the independent coefficients of mass matrices $M_S^2$, $M_P^2$,  $M_{S^\pm}^2$, ${M_n}$ and ${M_c}$ can be found in Ref.~\cite{ref-zhang1}.
Using $8\times8$ unitary matrices $R_S$, $R_P$ and $R_{S^\pm}$, the unrotated basises $S'$, $P'$ and ${S^\pm}'$ can be respectively rotated to the mass eigenvectors $S$, $P$ and ${S^\pm}$:
\begin{eqnarray}
&&\hspace{-1.0cm}S'=R_S S,\quad P'=R_P P,\quad {S^\pm}'=R_{S^\pm} {S^\pm}.
\end{eqnarray}
Through the unitary matrices $Z_n$, $Z_{-}$ and $Z_{+}$, neutral and charged fermions  can also be rotated to the mass eigenvectors $\chi^0$ and $\chi$, respectively.

\section{125 GeV Higgs decay with lepton flavor violation\label{sec3}}

\begin{figure}
\begin{center}
\begin{minipage}[c]{0.23\textwidth}
\includegraphics[width=1.8in]{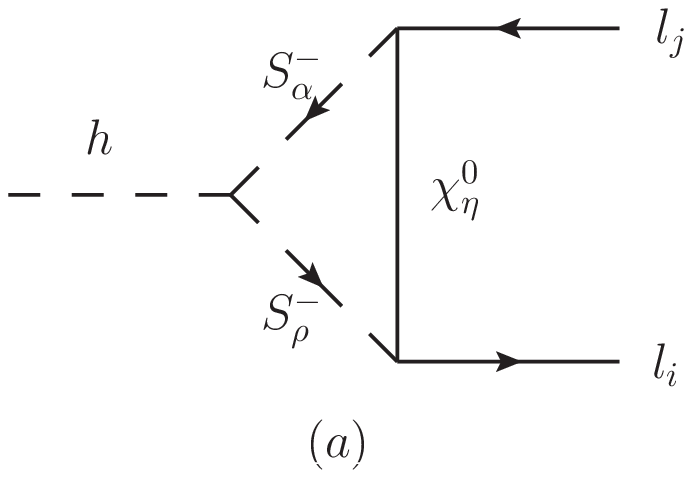}
\end{minipage}%
\begin{minipage}[c]{0.23\textwidth}
\includegraphics[width=1.8in]{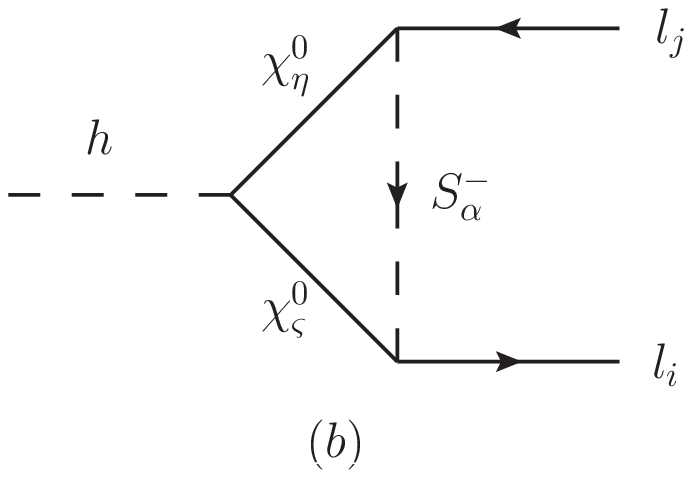}
\end{minipage}
\begin{minipage}[c]{0.23\textwidth}
\includegraphics[width=1.8in]{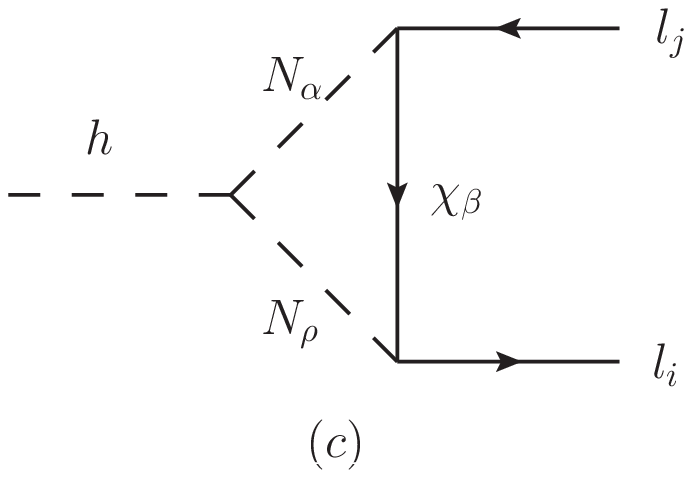}
\end{minipage}%
\begin{minipage}[c]{0.23\textwidth}
\includegraphics[width=1.8in]{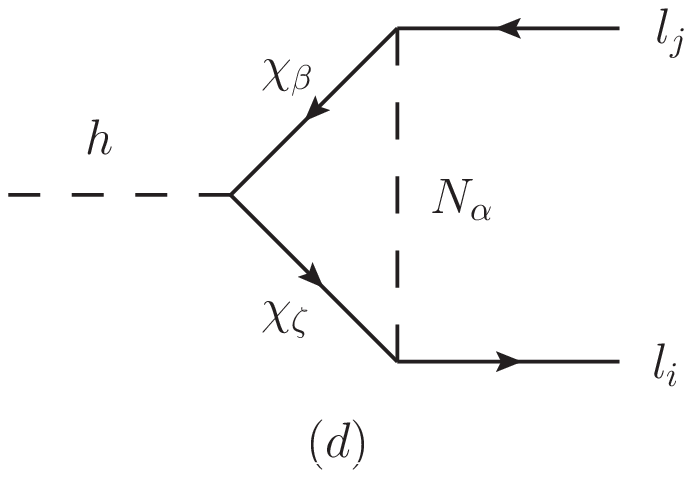}
\end{minipage}
\caption[]{\label{fig1} Vertex diagrams for $h\rightarrow \bar{l}_i l_j$. (a,b) represent the contributions from charged scalar $S_{\alpha,\rho}^-$ and neutral fermion $\chi_{\eta,\varsigma}^0$ loops, while (c,d) represent the contributions from neutral scalar $N_{\alpha,\rho}$ ($N=S,P$) and charged fermion $\chi_{\beta,\zeta}$ loops. }
\end{center}
\end{figure}

\begin{figure}
\begin{center}
\begin{minipage}[c]{0.23\textwidth}
\includegraphics[width=1.6in]{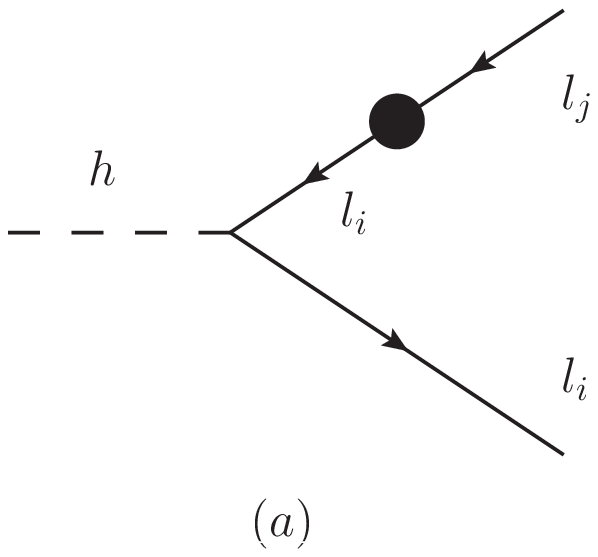}
\end{minipage}%
\begin{minipage}[c]{0.23\textwidth}
\includegraphics[width=1.6in]{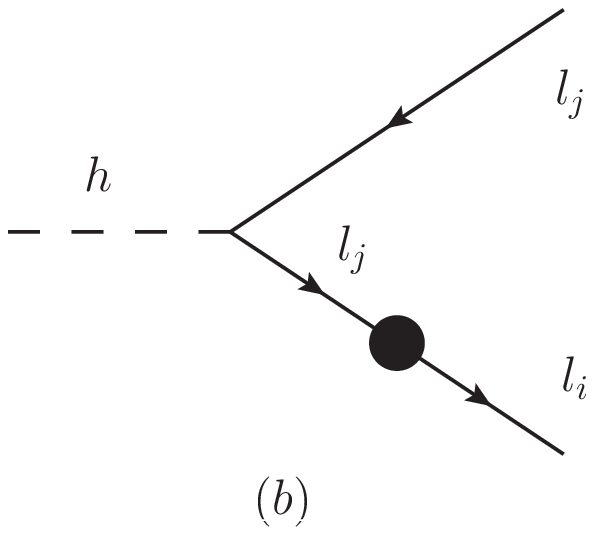}
\end{minipage}
\begin{minipage}[c]{0.23\textwidth}
\includegraphics[width=1.4in]{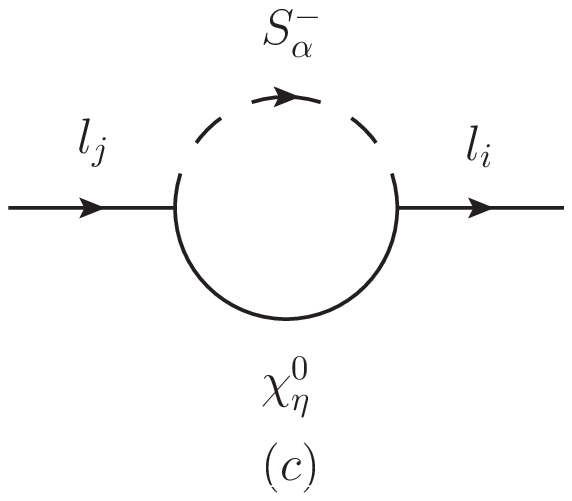}
\end{minipage}%
\begin{minipage}[c]{0.23\textwidth}
\includegraphics[width=1.4in]{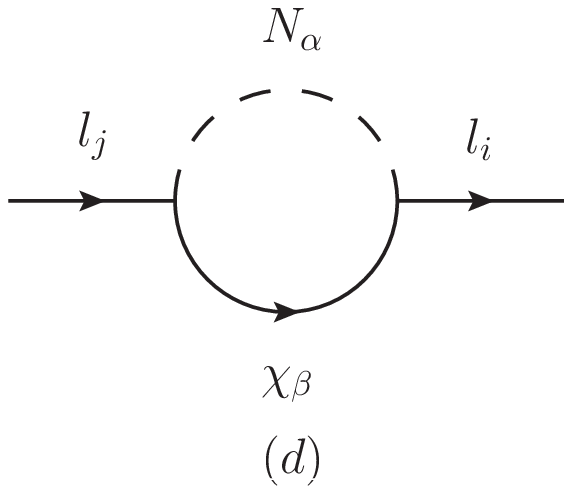}
\end{minipage}
\caption[]{\label{fig2-self} Self-energy diagrams contributing to $h\rightarrow \bar{l}_i l_j$ in the $\mu\nu$SSM. The blob in (a,b) indicates the self-energy contributions from (c,d). }
\end{center}
\end{figure}

The corresponding effective amplitude for 125 GeV Higgs decay with lepton flavor violation $h\rightarrow \bar{l}_i l_j$ can be written as
\begin{eqnarray}
\mathcal{M}= {\bar l_i}({F_L^{ij}}{P_L} + {F_R^{ij}}{P_R}){l_j},
\end{eqnarray}
with
\begin{eqnarray}
{F_{L,R}^{ij}} = F_{L,R}^{(V)ij} + F_{L,R}^{(S)ij},
\end{eqnarray}
where $F_{L,R}^{(V)ij}$ denotes the contributions from the vertex diagrams in Fig.~\ref{fig1}, and $F_{L,R}^{(S)ij}$ stands for the contributions from the self-energy diagrams in Fig.~\ref{fig2-self}, respectively.

The one-loop vertex diagrams for $h\rightarrow \bar{l}_i l_j$ in the $\mu\nu$SSM are depicted by Fig.~\ref{fig1}. Then, we can have
\begin{eqnarray}
F_{L,R}^{(V)ij} = F_{L,R}^{(a)ij} + F_{L,R}^{(b)ij} + F_{L,R}^{(c)ij} + F_{L,R}^{(d)ij},
\end{eqnarray}
where $F_{L,R}^{(a,b)ij}$ denotes the contributions from charged scalar $S_{\alpha,\rho}^-$ and neutral fermion $\chi_{\eta,\varsigma}^0$ loops, and $F_{L,R}^{(c,d)ij}$ stands for the contributions from the neutral scalar $N_{\alpha,\rho}$ ($N=S,P$) and charged fermion $\chi_{\beta,\zeta}$ loops, respectively. After integrating the heavy freedoms out, we formulate the neutral fermion loop contributions $F_{L,R}^{(a,b)ij}$ as follows:
\begin{eqnarray}
&&\hspace{-0.75cm}F_L^{(a)ij} =  \frac{{m_{{\chi _\eta ^0}}}{C^{S^\pm}_{1 \alpha \rho }}}{{m_W^2}}
C_L^{S_\rho ^ - \chi _\eta ^0{{\bar l }_{i}}}
C_L^{S_\alpha ^{-\ast} {l_{j}}\bar \chi _\eta ^0}
{G_1}({x_{\chi _\eta ^0}},{x_{S_\alpha ^ - }},{x_{S_\rho ^ - }}) ,\nonumber\\
&&\hspace{-0.75cm}F_L^{(b)ij} =  \frac{{m_{{\chi _\varsigma^0 }}}{m_{{\chi _\eta^0 }}}}{{m_W^2}}
C_L^{{S_\alpha^- }{\chi _\varsigma^0 }{{\bar l}_{i}}}C_L^{h{\chi _\eta^0 }{{\bar \chi }_\varsigma^0 }}
C_L^{{S_\alpha^{-\ast} }{l_{j}}{{\bar \chi }_\eta^0}}{G_1}({x_{{S_\alpha^- }}},{x_{{\chi _\varsigma^0 }}},{x_{{\chi _\eta^0 }}})\nonumber\\
&&\hspace{0.5cm} + \:  C_L^{{S_\alpha^- }{\chi _\varsigma^0 }{{\bar l}_{i}}}C_R^{h{\chi _\eta^0 }{{\bar \chi }_\varsigma^0 }}
C_L^{{S_\alpha^{-\ast} }{l_{j}}{{\bar \chi }_\eta^0}}{G_2}({x_{{S_\alpha^- }}},{x_{{\chi _\varsigma^0 }}},{x_{{\chi _\eta^0 }}}) ,\nonumber\\
&&\hspace{-0.75cm}F_R^{(a,b)ij} = \left. {F_L^{(a,b)ij}} \right|{ _{L \leftrightarrow R}} .
\end{eqnarray}
Here, the concrete expressions for couplings $C$ (and below) can be found in Appendix~\ref{app-coupling} and Ref.~\cite{ref-zhang2}, $x= {m^2}/{m_W^2}$, $m$ is the mass for the corresponding particle, and the loop functions $G_{i}$ are given as
\begin{eqnarray}
&&\hspace{-0.75cm}{G_1}({\textit{x}_1 , \textit{x}_2 , \textit{x}_3}) =  \frac{1}{{16{\pi ^2}}}\Big[ \frac{{{x_1}\ln {x_1}}}{{({x_2} - {x_1})({x_1} - {x_3})}}
+ \frac{{{x_2}\ln {x_2}}}{{({x_1} - {x_2})({x_2} - {x_3})}}  + \frac{{{x_3}\ln {x_3}}}{{({x_1} - {x_3})({x_3} - {x_2})}}\Big], \\
&&\hspace{-0.75cm}{G_2}({\textit{x}_1 , \textit{x}_2 , \textit{x}_3}) =  \frac{1}{{16{\pi ^2}}}\Big[  \frac{{x_1^2\ln {x_1}}}{{({x_2} - {x_1})({x_1} - {x_3})}}
+ \frac{{x_2^2\ln {x_2}}}{{({x_1} - {x_2})({x_2} - {x_3})}}  + \frac{{x_3^2\ln {x_3}}}{{({x_1} - {x_3})({x_3} - {x_2})}} \Big].\quad\;\;
\end{eqnarray}

In a similar way, the charged fermion loop contributions $F_{L,R}^{(c,d)ij}$ are
\begin{eqnarray}
&&\hspace{-0.75cm}F_L^{(c)ij} = \sum\limits_{N=S,P} \frac{{m_{{\chi _\beta }}}{C^{N}_{1 \alpha \rho }}}{{m_W^2}}
C_L^{N_\rho  \chi _\beta {{\bar l }_{i}}}
C_L^{N_\alpha  {l_{j}}\bar \chi _\beta }
{G_1}({x_{\chi _\beta }},{x_{N_\alpha  }},{x_{N_\rho  }}) ,\nonumber\\
&&\hspace{-0.75cm}F_L^{(d)ij} = \sum\limits_{N=S,P} \Big[ C_L^{{N_\alpha }{\chi _\zeta }{{\bar l}_{i}}}C_R^{h{\chi _\beta }{{\bar \chi }_\zeta }}C_L^{{N_\alpha }{l_{j}}{{\bar \chi }_\beta }}
{G_2}({x_{{N_\alpha }}},{x_{{\chi _\zeta }}},{x_{{\chi _\beta }}})\nonumber\\
&&\hspace{-0.1cm} + \frac{{m_{{\chi _\zeta }}}{m_{{\chi _\beta }}}}{{m_W^2}}
C_L^{{N_\alpha }{\chi _\zeta }{{\bar l}_{i}}}C_L^{h{\chi _\beta }{{\bar \chi }_\zeta }}
C_L^{{N_\alpha }{l_{j}}{{\bar \chi }_\beta }}{G_1}({x_{{N_\alpha }}},{x_{{\chi _\zeta }}},{x_{{\chi _\beta }}})  \Big],\nonumber\\
&&\hspace{-0.75cm}F_R^{(c,d)ij} = \left. {F_L^{(c,d)ij}} \right|{ _{L \leftrightarrow R}} .
\end{eqnarray}

In Fig.~\ref{fig2-self}, we show the self-energy diagrams contributing to $h\rightarrow \bar{l}_i l_j$ in the $\mu\nu$SSM. The contributions from the self-energy diagrams $F_{L,R}^{(S)ij}$ can be given as
\begin{eqnarray}
F_{L,R}^{(S)ij} = F_{L,R}^{(Sa)ij} + F_{L,R}^{(Sb)ij},
\end{eqnarray}
with
\begin{eqnarray}
&&\hspace{-0.75cm}F_{L}^{(Sa)ij} = \frac{C_L^{h l_i {\bar l}_i}}{m_{l_j}^2-m_{l_i}^2}
\Big\{ m_{l_j}^2 {\Sigma}_R (m_{l_j}^2) +  m_{l_j}^2 {\Sigma}_{Rs} (m_{l_j}^2)\nonumber\\
&&\hspace{0.7cm}
+\,  m_{l_i} [m_{l_j} {\Sigma}_L (m_{l_j}^2) + m_{l_j} {\Sigma}_{Ls} (m_{l_j}^2)] \Big\},\nonumber\\
&&\hspace{-0.75cm}F_{L}^{(Sb)ij} = \frac{C_L^{h l_j {\bar l}_j}}{m_{l_i}^2-m_{l_j}^2}
\Big\{ m_{l_i}^2 {\Sigma}_L (m_{l_i}^2) + m_{l_i}  m_{l_j} {\Sigma}_{Rs} (m_{l_i}^2) \nonumber\\
&&\hspace{0.7cm}
+\,  m_{l_j} [m_{l_i} {\Sigma}_R (m_{l_i}^2) + m_{l_j} {\Sigma}_{Ls} (m_{l_i}^2)] \Big\},\nonumber\\
&&\hspace{-0.75cm}F_{R}^{(Sa,Sb)ij} = \left. {F_{L}^{(Sa,Sb)ij}} \right|{ _{L \leftrightarrow R}}.
\end{eqnarray}
The ${\Sigma}$ of the self-energy diagrams in Fig.~\ref{fig2-self}(c,d) can be obtained below
\begin{eqnarray}
&&\hspace{-0.75cm}{\Sigma}_L (p^2) = -\frac{1}{16\pi^2}
\Big\{ B_1(p^2,m_{\chi_\eta^0}^2,m_{S_\alpha^-}^2)C_L^{S_\alpha ^ - \chi _\eta ^0{{\bar l }_{i}}}
C_R^{S_\alpha ^{-\ast} {l_{j}}\bar \chi _\eta ^0}\nonumber\\
&&\hspace{0.7cm}
+ \sum\limits_{N=S,P} B_1(p^2,m_{\chi_\beta}^2,m_{N_\alpha}^2)C_L^{{N_\alpha }{\chi _\beta }{{\bar l}_{i}}}C_R^{{N_\alpha }{l_{j}}{{\bar \chi }_\beta }}
 \Big\},\nonumber\\
&&\hspace{-0.75cm}m_{l_j} {\Sigma}_{Ls} (p^2) = \frac{1}{16\pi^2}
\Big\{ m_{\chi_\eta^0} B_0(p^2,m_{\chi_\eta^0}^2,m_{S_\alpha^-}^2)C_L^{S_\alpha ^ - \chi _\eta ^0{{\bar l }_{i}}}
C_L^{S_\alpha ^{-\ast} {l_{j}}\bar \chi _\eta ^0}\nonumber\\
&&\hspace{0.7cm}
+ \sum\limits_{N=S,P} m_{\chi_\beta} B_0(p^2,m_{\chi_\beta}^2,m_{N_\alpha}^2)C_L^{{N_\alpha }{\chi _\beta }{{\bar l}_{i}}}C_L^{{N_\alpha }{l_{j}}{{\bar \chi }_\beta }}
\Big\},\nonumber\\
&&\hspace{-0.75cm} {\Sigma}_{R} (p^2) = \left. {{\Sigma}_{L} (p^2)} \right|{ _{L \leftrightarrow R}},\nonumber\\
&&\hspace{-0.75cm} m_{l_j} {\Sigma}_{Rs} (p^2) = \left. {m_{l_j} {\Sigma}_{Ls} (p^2)} \right|{ _{L \leftrightarrow R}}.
\end{eqnarray}
Here, $B_{0,1}(p^2,m_0^2,m_1^2)$ are two-point functions~\cite{ref-B-0,ref-B-1,ref-B-2,ref-B-3,ref-B-4,ref-B-5,ref-B-6}.

Then, we can obtain the decay width of $h\rightarrow \bar{l}_i l_j$~\cite{LFVHD-Br1,LFVHD-Br2}
\begin{eqnarray}
{\Gamma}(h\rightarrow \bar{l}_i l_j) \simeq \frac{m_h}{16\pi}\Big({\left| {F_L^{ij}} \right|^2} + {\left| {F_R^{ij}} \right|^2}\Big).
\end{eqnarray}
If interpreted as a signal, the decay width of $h\rightarrow l_i l_j$ is
\begin{eqnarray}
\Gamma(h\rightarrow l_i l_j)= {\Gamma}(h\rightarrow \bar{l}_i l_j)+{\Gamma}(h\rightarrow \bar{l}_j l_i ),
\end{eqnarray}
and the branching ratio of $h\rightarrow l_i l_j$ is
\begin{eqnarray}
{\rm{Br}}(h\rightarrow l_i l_j)= {\Gamma(h\rightarrow l_i l_j)}/{{\Gamma}_h},
\end{eqnarray}
where ${{\Gamma}}_h \simeq 4.1\times10^{-3}\:{\rm{GeV}}$~\cite{HCSWG} denotes the total decay width of the 125 GeV Higgs boson.

\section{Numerical analysis\label{sec4}}
In order to obtain transparent numerical results in the $\mu\nu{\rm SSM}$, we take the minimal flavor violation (MFV) assumptions for some parameters, which assume
\begin{eqnarray}
&&\hspace{-0.9cm}{\kappa _{ijk}} = \kappa {\delta _{ij}}{\delta _{jk}}, \quad
{({A_\kappa }\kappa )_{ijk}} =
{A_\kappa }\kappa {\delta _{ij}}{\delta _{jk}}, \quad
\lambda _i = \lambda , \nonumber\\
&&\hspace{-0.9cm}
{{\rm{(}}{A_\lambda }\lambda {\rm{)}}_i} = {A_\lambda }\lambda,\quad
{Y_{{e_{ij}}}} = {Y_{{e_i}}}{\delta _{ij}},\quad
{Y_{{\nu _{ij}}}} = {Y_{{\nu _i}}}{\delta _{ij}},\nonumber\\
&&\hspace{-0.9cm}
\upsilon_{\nu_i^c}=\upsilon_{\nu^c},\quad
(A_\nu Y_\nu)_{ij}={a_{{\nu_i}}}{\delta _{ij}},\quad m_{\tilde \nu_{ij}^c}^2 = m_{{{\tilde \nu_i}^c}}^2{\delta _{ij}},
\nonumber\\
&&\hspace{-0.9cm}m_{\tilde Q_{ij}}^2 = m_{{{\tilde Q_i}}}^2{\delta _{ij}}, \quad
m_{\tilde u_{ij}^c}^2 = m_{{{\tilde u_i}^c}}^2{\delta _{ij}}, \quad
m_{\tilde d_{ij}^c}^2 = m_{{{\tilde d_i}^c}}^2{\delta _{ij}},
\label{MFV}
\end{eqnarray}
where $i,\;j,\;k =1,\;2,\;3 $. $m_{\tilde \nu_i^c}^2$ can be constrained by the minimization conditions of the neutral scalar potential seen in Ref.~\cite{ref-zhang1}. To agree with experimental observations on quark mixing, one can have
\begin{eqnarray}
&&\hspace{-0.75cm}\;\,{Y_{{u _{ij}}}} = {Y_{{u _i}}}{V_{L_{ij}}^u},\quad
 (A_u Y_u)_{ij}={A_{u_i}}{Y_{{u_{ij}}}},\nonumber\\
&&\hspace{-0.75cm}\;\,{Y_{{d_{ij}}}} = {Y_{{d_i}}}{V_{L_{ij}}^d},\quad
(A_d Y_d)_{ij}={A_{d_i}}{Y_{{d_{ij}}}},
\end{eqnarray}
and $V=V_L^u V_L^{d\dag}$ denotes the CKM matrix.

For the trilinear coupling matrix $({A_e}{Y_e})$ and soft breaking slepton mass matrices $m_{{{\tilde L},{\tilde e^c}}}^2$, we will take into account the off-diagonal terms for the matrices, which are named the slepton flavor mixings and are defined by~\cite{sl-mix,sl-mix1,sl-mix2,sl-mix3,sl-mix4,neu-zhang2}
\begin{eqnarray}
&&\hspace{-0.75cm}\quad\;\,{m_{\tilde L}^2} = \left( {\begin{array}{*{20}{c}}
   1 & \delta_{12}^{LL} & \delta_{13}^{LL}  \\
   \delta_{12}^{LL} & 1 & \delta_{23}^{LL}  \\
   \delta_{13}^{LL} & \delta_{23}^{LL} & 1  \\
\end{array}} \right){m_{L}^2},\\
&&\hspace{-0.75cm}\quad\:{m_{\tilde e^c}^2} = \left( {\begin{array}{*{20}{c}}
   1 & \delta_{12}^{RR} & \delta_{13}^{RR}  \\
   \delta_{12}^{RR} & 1 & \delta_{23}^{RR}  \\
   \delta_{13}^{RR} & \delta_{23}^{RR} & 1  \\
\end{array}} \right){m_{E}^2},\\
&&\hspace{-0.75cm}({A_e}{Y_e}) = \left( {\begin{array}{*{20}{c}}
   m_{l_1}{A_e} & \delta_{12}^{LR}{m_{L}}{m_{E}} & \delta_{13}^{LR}{m_{L}}{m_{E}}  \\
   \delta_{12}^{LR}{m_{L}}{m_{E}} & m_{l_2}{A_e} & \delta_{23}^{LR}{m_{L}}{m_{E}}  \\
   \delta_{13}^{LR}{m_{L}}{m_{E}} & \delta_{23}^{LR}{m_{L}}{m_{E}} & m_{l_3}{A_e}  \\
\end{array}} \right){1\over {\upsilon_d}}.
\end{eqnarray}
The following numerical results will show that the branching ratio of $h\rightarrow \mu \tau$ depends on the slepton mixing parameters $\delta_{23}^{XX}~(X=L,R)$.

At first, the constraints from some experiments should be considered. Through our previous work~\cite{neu-zhang1}, we have discussed in detail how the neutrino oscillation data constrain neutrino Yukawa couplings $Y_{\nu_i} \sim \mathcal{O}(10^{-7})$ and left-handed sneutrino VEVs $\upsilon_{\nu_i} \sim \mathcal{O}(10^{-4}{\rm{GeV}})$  via the seesaw mechanism. Here, due to the neutrino sector only weakly affecting $h\rightarrow \mu \tau$, we can take no account of the constraints from neutrino experiment data.

The neutral Higgs with mass around $125\;{\rm GeV}$ reported by ATLAS and CMS contributes a strict constraint on the relevant parameters of the $\mu\nu{\rm SSM}$. For a large mass of the pseudoscalar $M_A$ and moderate $\tan\beta$, the SM-like Higgs mass of the $\mu\nu{\rm SSM}$ is approximately written as~\cite{mnSSM1,ref-zhang3}
\begin{eqnarray}
m_h^2 \simeq m_Z^2 \cos^2 2\beta + \frac{6 \lambda^2 s_{_W}^2 c_{_W}^2}{ e^2} m_Z^2 \sin^2 2\beta+\bigtriangleup m_h^2.
\label{eq-mh}
\end{eqnarray}
Compared with the MSSM, the $\mu\nu{\rm SSM}$ gets an additional term, $\frac{6 \lambda^2 s_{_W}^2 c_{_W}^2}{ e^2} m_Z^2 \sin^2 2\beta$. Thus, the SM-like Higgs in the $\mu\nu{\rm SSM}$ can easily account for the mass around $125\,{\rm GeV}$, especially for small $\tan\beta$.
Including two-loop leading-log effects, the main radiative corrections $\bigtriangleup m_h^2$ can be given as~\cite{ref-mh-rad1,ref-mh-rad2,ref-mh-rad3}
\begin{eqnarray}
\bigtriangleup m_h^2 = \frac{3m_t^4}{4\pi^2\upsilon^2}\Big[(t+\frac{1}{2}{\tilde X_t})+\frac{1}{16\pi^2} (\frac{3m_t^2}{2\upsilon^2}-32\pi \alpha_3) (t^2+{\tilde X_t}t) \Big],\nonumber
\end{eqnarray}
\begin{eqnarray}
t=\log \frac{M_S^2}{m_t^2},\qquad{\tilde X_t}= \frac{2\tilde A_t^2}{M_S^2} \Big(1-\frac{\tilde A_t^2}{12M_S^2} \Big),
\end{eqnarray}
where $\upsilon=174$ GeV, $\alpha_3$ is the strong coupling constant, $M_S =\sqrt{m_{{\tilde t}_1}m_{{\tilde t}_2}}$ with $m_{{\tilde t}_{1,2}}$ denoting the stop masses, $\tilde A_t = A_t-\mu\cot\beta$ with $A_t=A_{u_{3}}$ being the trilinear Higgs-stop coupling and $\mu=3\lambda \upsilon_{\nu^c}$ denoting the Higgsino mass parameter.

We also impose a constraint on the SUSY contribution to the muon magnetic dipole moment $a_\mu$ in the $\mu\nu$SSM, which is given in Appendix~\ref{app-MDM} for convenience. The difference between experiment and the SM prediction on $a_\mu$ is~\cite{PDG,E821-1,E821-2}
\begin{eqnarray}
\Delta a_\mu =a_\mu^{{\rm{exp}}} -a_\mu^{{\rm{SM}}} = (24.8\pm7.9)\times 10^{-10},
\label{MDM-exp}
\end{eqnarray}
with all errors combining in quadrature. Therefore, the SUSY contribution to $a_\mu$ in the $\mu\nu$SSM should be constrained as $1.1\times 10^{-10} \leq \Delta a_\mu \leq 48.5\times 10^{-10}$, where a $3 \sigma$ experimental error is considered.

Through analysis of the parameter space of the $\mu\nu$SSM in Ref.~\cite{mnSSM1}, we can take reasonable parameter values to  be $\lambda=0.1$, $\kappa=0.4$, $A_\lambda=500\;{\rm GeV}$, ${A_{\kappa}}=-300\;{\rm GeV}$ and  ${A_{e}}=1\;{\rm TeV}$ for simplicity. For the gauginos' Majorana masses, we will choose the approximate GUT relation $M_1=\frac{\alpha_1^2}{\alpha_2^2}M_2\approx 0.5 M_2$ and $M_3=\frac{\alpha_3^2}{\alpha_2^2}M_2\approx 2.7 M_2$. The gluino mass, $m_{{\tilde g}}\approx M_3$, is greater than about $1.2$ TeV from the ATLAS and CMS experimental data~\cite{ATLAS-sg1,ATLAS-sg2,CMS-sg1,CMS-sg2}. For simplicity, we could adopt $m_{{\tilde Q}_3}=m_{{\tilde u}^c_3}=m_{{\tilde d}^c_3}=1.5\;{\rm TeV}$. As key parameters, $A_t$ and $\tan\beta \equiv \upsilon_u/\upsilon_d$ affect the SM-like Higgs mass. Here, we keep the SM-like Higgs mass $m_h=125\;{\rm GeV}$ as input, and then the value of parameter $A_t$ can be given automatically in the numerical calculation. Then, the free parameters that affect our next analysis are $\tan \beta ,\;\mu\equiv3\lambda \upsilon_{\nu^c},\; M_2,\; m_{L},\; m_{E}$ and slepton mixing parameters $\delta_{23}^{XX}~(X=L,R)$.

\begin{table*}
\begin{tabular*}{\textwidth}{@{\extracolsep{\fill}}lllll@{}}
\hline
Parameters&Min&Max&Step\\
\hline
$\tan \beta$&5&50&15\\
$\mu=M_2/{\rm GeV}$&500&5000&500\\
$m_{L}=m_{E}/{\rm GeV}$&500&5000&500\\
\hline
$\delta_{23}^{LR}$&0&0.4&0.02\\
$\delta_{23}^{LL}$&0&1.0&0.05\\
$\delta_{23}^{RR}$&0&1.0&0.05\\
\hline
\end{tabular*}
\caption{Scanning parameters for Fig.~\ref{fig-dLR}.}
\label{tab1}
\end{table*}

\begin{figure}
\begin{center}
\begin{minipage}[c]{0.48\textwidth}
\includegraphics[width=2.6in]{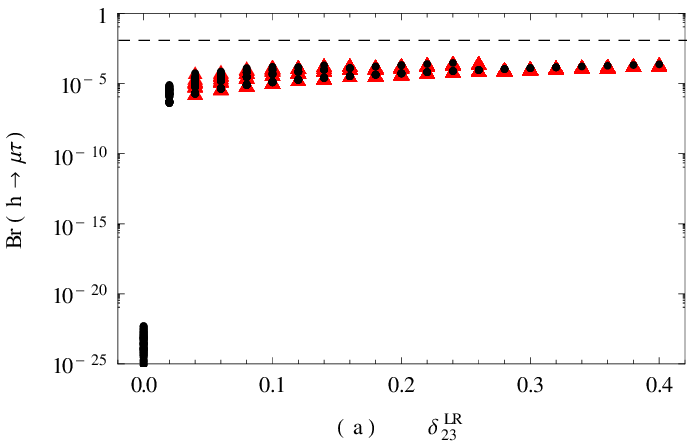}
\end{minipage}%
\begin{minipage}[c]{0.45\textwidth}
\includegraphics[width=2.6in]{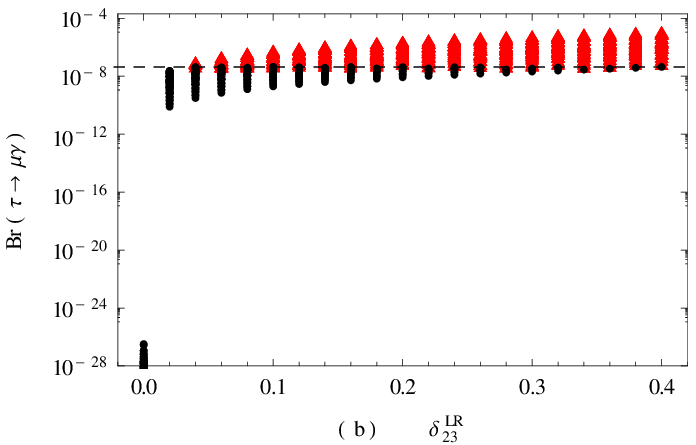}
\end{minipage}
\begin{minipage}[c]{0.48\textwidth}
\includegraphics[width=2.6in]{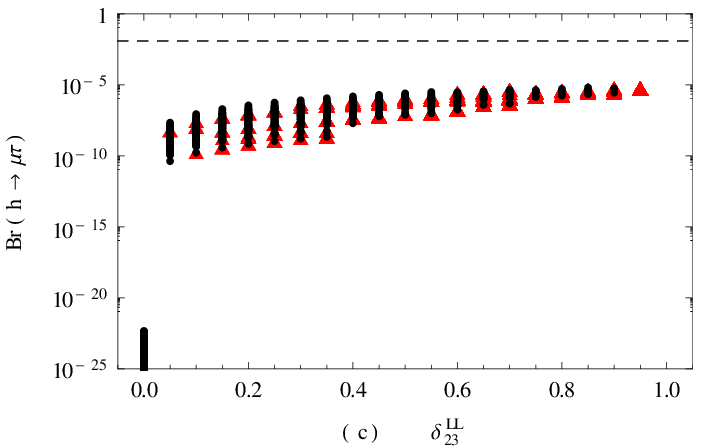}
\end{minipage}%
\begin{minipage}[c]{0.45\textwidth}
\includegraphics[width=2.6in]{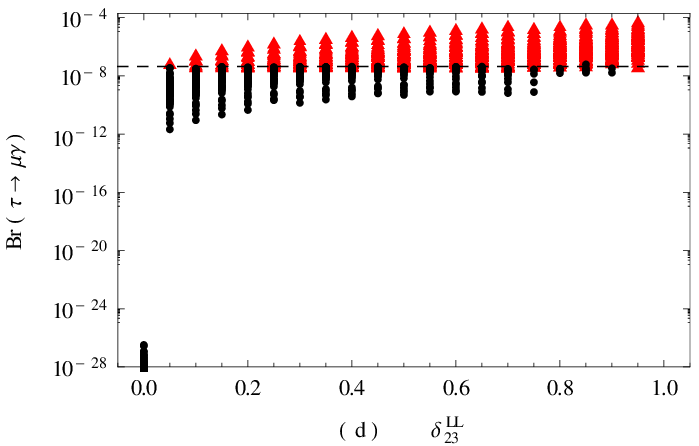}
\end{minipage}
\begin{minipage}[c]{0.48\textwidth}
\includegraphics[width=2.6in]{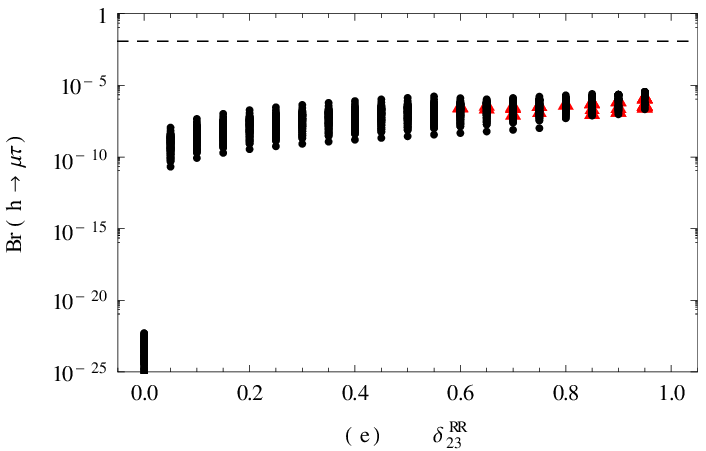}
\end{minipage}%
\begin{minipage}[c]{0.45\textwidth}
\includegraphics[width=2.6in]{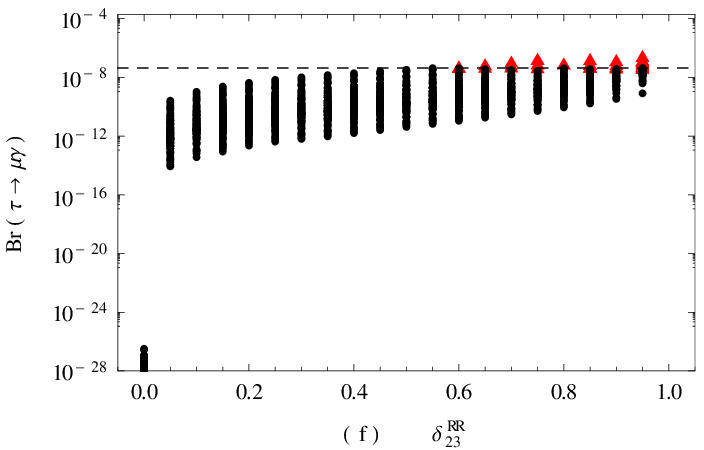}
\end{minipage}
\caption[]{\label{fig-dLR}  (Color online) ${\rm{Br}}(h\rightarrow \mu \tau)$  versus slepton mixing parameters $\delta_{23}^{LR}$ (a), $\delta_{23}^{LL}$ (c), and $\delta_{23}^{RR}$ (e), where the dashed line stands for the upper limit on ${\rm{Br}}(h\rightarrow \mu \tau)$ at 95\% CL showed in Eq.~(\ref{uplimit}).  ${\rm{Br}}(\tau \rightarrow \mu \gamma)$ versus slepton mixing parameters $\delta_{23}^{LR}$ (b), $\delta_{23}^{LL}$ (d), and $\delta_{23}^{RR}$ (f), where the dashed line denotes the present limit of ${\rm{Br}}(\tau\rightarrow \mu\gamma)$ seen in Eq.~(\ref{up-tur}). Here, the red triangles are ruled out by the present limit of ${\rm{Br}}(\tau\rightarrow \mu\gamma)$, and the black circles are consistent with the present limit of ${\rm{Br}}(\tau\rightarrow \mu\gamma)$.}
\end{center}
\end{figure}

It is well known that the lepton flavour violating processes are flavor dependent. The LFV rates for $\mu-\tau$  transitions depend on the slepton mixing parameters $\delta_{23}^{XX}~(X=L,R)$, which can be confirmed by Fig.~\ref{fig-dLR}. The slepton mixing parameters $\delta_{12}^{XX}$ and $\delta_{13}^{XX}~(X=L,R)$ hardly affect the LFV rates for $\mu-\tau$  transitions, which play a leading role in the LFV rates for $e-\mu$ and $e-\tau$ transitions. So, we take  $\delta_{12}^{XX}=0$ and $\delta_{13}^{XX}=0~(X=L,R)$ here. To produce Fig.~\ref{fig-dLR}, we scan the parameter space shown in Tab.~\ref{tab1}. Here the steps are large, because the running of the program is not very fast. However the scanned parameter space is broad enough  to contain the possibility of more.

In the scan, we keep the chargino masses $m_{\chi_\beta}> 200$ GeV $(\beta=1,2)$, the neutral fermion masses $m_{\chi_\eta^0}> 200$ GeV $(\eta=1,\cdots,7)$, and the scalar masses $m_{S_\alpha,P_\alpha,S_\alpha^\pm}> 500$ GeV $(\eta=2,\cdots,8)$, to avoid the range ruled out by the experiments~\cite{PDG}. The results are also constrained by the muon anomalous magnetic dipole moment $1.1\times 10^{-10} \leq \Delta a_\mu \leq 48.5\times 10^{-10}$, where a $3 \sigma$ experimental error is considered. In Ref.~\cite{ref-zhang2}, we have investigated the signals of the Higgs boson decay channels $h\rightarrow\gamma\gamma$, $h\rightarrow VV^*$ ($V=Z,W$), and $h\rightarrow f\bar{f}$ ($f=b,\tau$) in the $\mu\nu$SSM. When the lightest stop mass $m_{{\tilde t}_1}\gtrsim 700\;{\rm GeV}$ and the lightest stau mass $m_{{\tilde \tau}_1}\gtrsim 300\;{\rm GeV}$, the signal strengths of these Higgs boson decay channels are in agreement with the SM. Therefore, the scanning results in this paper coincide with the experimental data of these Higgs boson decay channels.

Note that, when the calculation program is scanning one of the slepton mixing parameters $\delta_{23}^{XX}~(X=L,R)$, the other two slepton mixing parameters $\delta_{23}^{XX}~(X=L,R)$ are set to zero. So, we can see the contribution of every slepton mixing parameter alone. Then in Fig.~\ref{fig-dLR}, we plot ${\rm{Br}}(h\rightarrow \mu \tau)$  varying with slepton mixing parameters $\delta_{23}^{LR}$ (a), $\delta_{23}^{LL}$ (c), and $\delta_{23}^{RR}$ (e) respectively, where the dashed line stands for the upper limit on ${\rm{Br}}(h\rightarrow \mu \tau)$ at 95\% CL shown in Eq.~(\ref{uplimit}). We also plot ${\rm{Br}}(\tau \rightarrow \mu \gamma)$ versus slepton mixing parameters $\delta_{23}^{LR}$ (b), $\delta_{23}^{LL}$ (d), and $\delta_{23}^{RR}$ (f) respectively, where the dashed line denotes the present limit of ${\rm{Br}}(\tau\rightarrow \mu\gamma)$~\cite{t-exp1}:
\begin{eqnarray}
{\rm{Br}}(\tau\rightarrow \mu\gamma)<4.4\times10^{-8}.
\label{up-tur}
\end{eqnarray}
Here, the red triangles are ruled out by the present limit of ${\rm{Br}}(\tau\rightarrow \mu\gamma)$, and the black circles are consistent with the present limit of ${\rm{Br}}(\tau\rightarrow \mu\gamma)$.

In Fig.~\ref{fig-dLR}, when slepton mixing parameters $\delta_{23}^{XX}=0~(X=L,R)$,  ${\rm{Br}}(h\rightarrow \mu \tau)$ can reach $\mathcal{O}(10^{-23})$ and ${\rm{Br}}(\tau\rightarrow \mu\gamma)$ can attain $\mathcal{O}(10^{-27})$, because the contributions can come from the mixing of the particles, which can  easily be seen in Eq.~(\ref{eq-quad}). These results are too small to detect. However, if the nonzero slepton mixing parameters $\delta_{23}^{XX}~(X=L,R)$ are considered, ${\rm{Br}}(h\rightarrow \mu \tau)$ and ${\rm{Br}}(\tau\rightarrow \mu\gamma)$ grow quickly. With increasing  $\delta_{23}^{XX}~(X=L,R)$, ${\rm{Br}}(\tau\rightarrow \mu\gamma)$ can easily go beyond the present experimental limit of ${\rm{Br}}(\tau\rightarrow \mu\gamma)$, shown in the plot as the red triangles. Although ${\rm{Br}}(h\rightarrow \mu \tau)$ cannot reach the present experimental upper limit of ${\rm{Br}}(h\rightarrow \mu \tau)$, ${\rm{Br}}(h\rightarrow \mu \tau)$ becomes larger and approaches the present experimental limit with increasing  $\delta_{23}^{XX}~(X=L,R)$. Especially in Fig.~\ref{fig-dLR}(a), considering  nonzero slepton mixing parameters $\delta_{23}^{LR}$, ${\rm{Br}}(h\rightarrow \mu \tau)$ can achieve $\mathcal{O}(10^{-4})$, which is below the present experimental limit by just two orders of magnitude. Compared to the MSSM, exotic singlet righthanded neutrino superfields in the $\mu\nu$SSM induce new sources for lepton-flavor violation, considering that the righthanded neutrino and sneutrinos can mix and couple with the other particles seen in Eq.~(\ref{eq-quad}) and Appendix~\ref{app-coupling}. In Fig.~\ref{fig-dLR}(a,c,e), the red triangles overlap with the black circles, because some parameters strongly affect ${\rm{Br}}(\tau\rightarrow \mu\gamma)$ but do not affect ${\rm{Br}}(h\rightarrow \mu \tau)$. We will research this further in the following.

\begin{table*}
\begin{tabular*}{\textwidth}{@{\extracolsep{\fill}}lllll@{}}
\hline
Parameters&Min&Max&Step\\
\hline
$\tan \beta$&5&50&2.5\\
$M_{\rm{SUSY}}/{\rm GeV}$&500&5000&250\\
\hline
\end{tabular*}
\caption{Scanning parameters for Fig.~\ref{fig-tbm}, where $\mu=M_2=m_{L}=m_{E}\equiv M_{\rm{SUSY}}$.}
\label{tab2}
\end{table*}

\begin{figure}
\begin{center}
\begin{minipage}[c]{0.48\textwidth}
\includegraphics[width=2.6in]{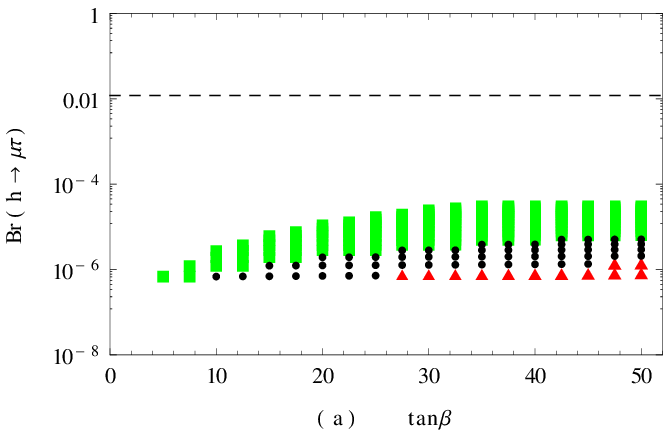}
\end{minipage}%
\begin{minipage}[c]{0.45\textwidth}
\includegraphics[width=2.6in]{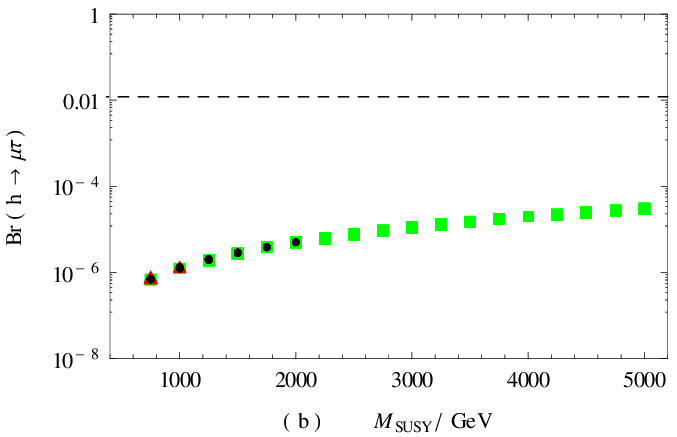}
\end{minipage}
\begin{minipage}[c]{0.48\textwidth}
\includegraphics[width=2.6in]{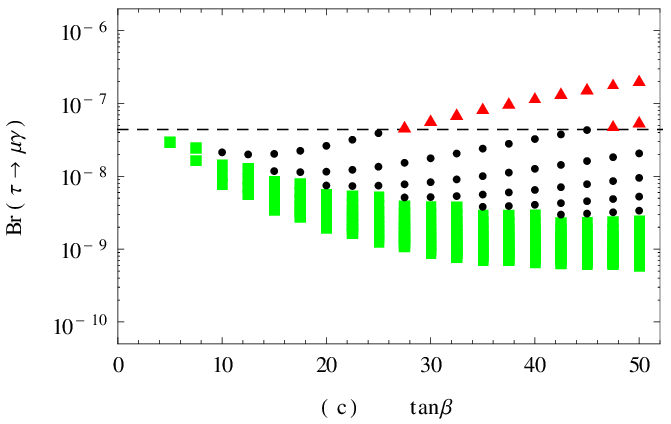}
\end{minipage}%
\begin{minipage}[c]{0.45\textwidth}
\includegraphics[width=2.6in]{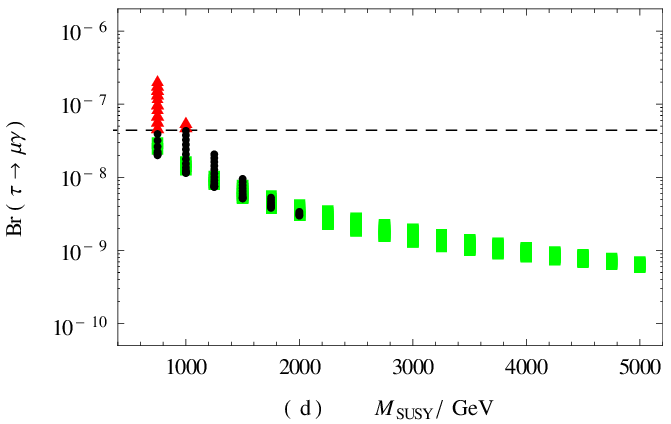}
\end{minipage}
\begin{minipage}[c]{0.48\textwidth}
\includegraphics[width=2.6in]{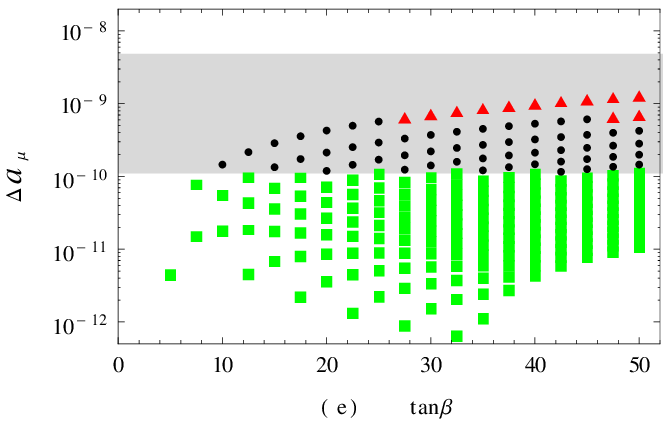}
\end{minipage}%
\begin{minipage}[c]{0.45\textwidth}
\includegraphics[width=2.6in]{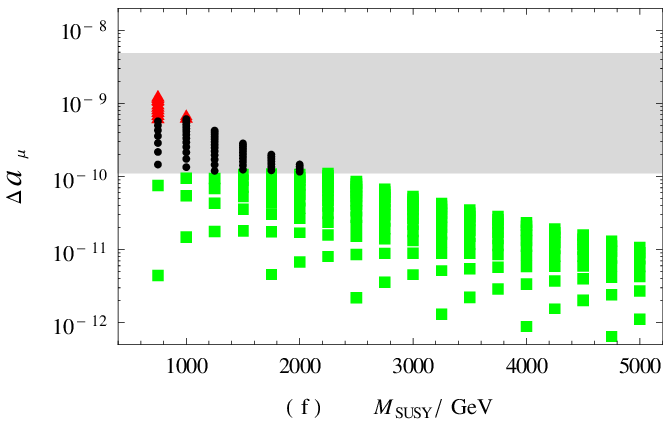}
\end{minipage}
\caption[]{\label{fig-tbm} (Color online) ${\rm{Br}}(h\rightarrow \mu \tau)$  versus $\tan \beta$ (a) and $M_{\rm{SUSY}}$ (b), where the dashed line stands for the upper limit on ${\rm{Br}}(h\rightarrow \mu \tau)$ at 95\% CL shown in Eq.~(\ref{uplimit}).  ${\rm{Br}}(\tau \rightarrow \mu \gamma)$ versus $\tan \beta$ (c) and $M_{\rm{SUSY}}$ (d), where the dashed line denotes the present limit of ${\rm{Br}}(\tau\rightarrow \mu\gamma)$, which can be seen in Eq.~(\ref{up-tur}).  $\Delta a_\mu$ versus $\tan \beta$ (e) and $M_{\rm{SUSY}}$ (f), where the gray area denotes the $\Delta a_\mu$ at $3.0\sigma$ given in Eq.~(\ref{MDM-exp}). Here, the red triangles are excluded by the present limit of ${\rm{Br}}(\tau\rightarrow \mu\gamma)$, the green squares are eliminated by the $\Delta a_\mu$ at $3.0\sigma$, and the black circles simultaneously conform to the present limit of ${\rm{Br}}(\tau\rightarrow \mu\gamma)$ and the $\Delta a_\mu$ at $3.0\sigma$. }
\end{center}
\end{figure}
To see how other parameters affect the results, we appropriately fix $\delta_{23}^{LR}=0.02$ and $\delta_{23}^{LL}=\delta_{23}^{RR}=0.2$. Then, we scan the parameter space shown in Table~\ref{tab2}, where $\mu=M_2=m_{L}=m_{E}\equiv M_{\rm{SUSY}}$. In the scanning, we also keep the chargino masses $m_{\chi_\beta}> 200$ GeV $(\beta=1,2)$, the neutral fermion masses $m_{\chi_\eta^0}> 200$ GeV $(\eta=1,\cdots,7)$, and the scalar masses $m_{S_\alpha,P_\alpha,S_\alpha^\pm}> 500$ GeV $(\eta=2,\cdots,8)$, to avoid the range ruled out by the experiments~\cite{PDG}. Then in Fig.~\ref{fig-tbm}, we plot ${\rm{Br}}(h\rightarrow \mu \tau)$ respectively versus $\tan \beta$ (a) and $M_{\rm{SUSY}}$ (b), where the dashed line stands for the upper limit on ${\rm{Br}}(h\rightarrow \mu \tau)$ at 95\% CL shown in Eq.~(\ref{uplimit}). We show ${\rm{Br}}(\tau \rightarrow \mu \gamma)$ varying with $\tan \beta$ (c) and $M_{\rm{SUSY}}$ (d) respectively, where the dashed line denotes the present limit of ${\rm{Br}}(\tau\rightarrow \mu\gamma)$ which can be seen in Eq.~(\ref{up-tur}). We also picture the muon anomalous magnetic dipole moment $\Delta a_\mu$  versus $\tan \beta$ (e) and $M_{\rm{SUSY}}$ (f) respectively, where the gray area denotes the $\Delta a_\mu$ at $3.0\sigma$ given in Eq.~(\ref{MDM-exp}). Here, the red triangles are excluded by the present limit of ${\rm{Br}}(\tau\rightarrow \mu\gamma)$, the green squares are eliminated by the $\Delta a_\mu$ at $3.0\sigma$, and the black circles conform to both the present limit of ${\rm{Br}}(\tau\rightarrow \mu\gamma)$ and the  $\Delta a_\mu$ at $3.0\sigma$.

In Fig.~\ref{fig-tbm}(d,f), the numerical results show that ${\rm{Br}}(\tau\rightarrow \mu\gamma)$ and the muon anomalous magnetic dipole moment $\Delta a_\mu$ are decoupling with increasing $M_{\rm{SUSY}}$. For large $M_{\rm{SUSY}}$, it is hard to give large contribution to $\Delta a_\mu$. So, the large $M_{\rm{SUSY}}$ are easily excluded by the $\Delta a_\mu$ at $3.0\sigma$ given in Eq.~(\ref{MDM-exp}), which can be seen in the graph as the green squares. For small $M_{\rm{SUSY}}$, there can be a large contribution to ${\rm{Br}}(\tau\rightarrow \mu\gamma)$. Therefore, the small $M_{\rm{SUSY}}$ are easily ruled out by the present experimental limit of ${\rm{Br}}(\tau\rightarrow \mu\gamma)$, shown as the red triangles. In Fig.~\ref{fig-tbm}(b), ${\rm{Br}}(h\rightarrow \mu \tau)$ is non-decoupling with increasing  $M_{\rm{SUSY}}$, which is in agreement with the research in the MSSM~\cite{LFVHD23,LFVHD43}. Due to the introduction of slepton mixing parameters, the non-decoupling behaviour of ${\rm{Br}}(h\rightarrow \mu \tau)$ tends to $\mathcal{O}((m_h/M_{SUSY})^{0})$, which is somewhat different from the Appelquist-Carazzone decoupling theorem~\cite{appelquist}. (As a side note, in Ref.~\cite{Draper-Haber}, a non-decoupling behaviour in computation of the Higgs mass showed that it was linked to an ambiguity in the treatment of $\tan\beta$, which is a renormalization scheme dependent parameter.) We can also see that the red triangles overlap with the black circles in Fig.~\ref{fig-tbm}(b), because the parameter $\tan \beta$ does not affect ${\rm{Br}}(h\rightarrow \mu \tau)$ visibly in this parameter space. In Fig.~\ref{fig-tbm}(a,c,e), the numerical results show that ${\rm{Br}}(h\rightarrow \mu \tau)$, ${\rm{Br}}(\tau\rightarrow \mu\gamma)$ and the muon anomalous magnetic dipole moment $\Delta a_\mu$ can have large values when $\tan \beta$ is large.

\section{Summary\label{sec5}}
In this work, we have studied the 125 GeV Higgs decay with lepton flavor violation, $h\rightarrow \mu \tau$, in the framework of the $\mu\nu$SSM with slepton flavor mixing. The numerical results show that the branching ratio of $h\rightarrow \mu \tau$ depends on the slepton mixing parameters $\delta_{23}^{XX}~(X=L,R)$, because the lepton flavour violating  processes are flavor dependent. The branching ratio of $h\rightarrow \mu \tau$ increases with increasing $\delta_{23}^{XX}~(X=L,R)$. Under the experimental constraints of the muon anomalous magnetic dipole moment, the SM-like Higgs mass around 125 GeV and the present limit of ${\rm{Br}}(\tau\rightarrow \mu\gamma)$, the branching ratio of $h\rightarrow \mu \tau$ can reach $\mathcal{O}(10^{-4})$. Compared with the MSSM, exotic singlet righthanded neutrino superfields in the $\mu\nu$SSM induce new sources for the lepton-flavor violation. Considering that the recent ATLAS and CMS measurements for $h\rightarrow \mu \tau$ do not show a significant deviation from the SM, the experiments still need to make more precise measurements in the future. To detect a Higgs boson lepton flavour violating process is a prospective window to search for new physics.

\begin{acknowledgments}
Supported by Major Project of NNSFC (11535002) and NNSFC (11275036, 11647120),
the Open Project Program of State Key Laboratory of Theoretical Physics,
Institute of Theoretical Physics, Chinese Academy of Sciences, China (Y5KF131CJ1),
the Natural Science Foundation of Hebei province (A2013201277, A2016201010, A2016201069),
Hebei Key Lab of Optic-Electronic Information and Materials, Midwest Universities Comprehensive Strength Promotion Project.
\end{acknowledgments}

\appendix

\section{The couplings\label{app-coupling}}

The couplings between CP-even neutral scalars and the other CP-even (or CP-odd) neutral scalars are formulated as
\begin{eqnarray}
\mathcal{L}_{int} =  C_{\alpha \beta \gamma}^{S} S_\alpha S_\beta S_\gamma + C_{\alpha \beta \gamma}^{P} S_\alpha P_\beta P_\gamma ,
\end{eqnarray}
with
\begin{eqnarray}
&&C_{\alpha \beta \gamma}^{S} = \frac{-e^2}{4\sqrt{2}{s_{_W}^2}{c_{_W}^2}} \Big[ \upsilon_d R_S^{1\alpha}R_{S}^{1\beta}R_{S}^{1\gamma}
+ \upsilon_u R_S^{2\alpha}R_{S}^{2\beta}R_{S}^{2\gamma}
+(\upsilon_d R_S^{1\alpha} + \upsilon_u R_S^{2\alpha}) R_{S}^{(2+i)\beta}R_{S}^{(2+i)\gamma} \Big]
 \nonumber\\
&&\hspace{1.2cm}
+ \frac{1}{\sqrt{2}} \Big[ \lambda_i \lambda_i (\upsilon_d R_S^{1\alpha}R_{S}^{2\beta}R_{S}^{2\gamma}
+ \upsilon_u R_S^{2\alpha}R_{S}^{1\beta}R_{S}^{1\gamma})
- \lambda_i \lambda_j (\upsilon_d R_S^{1\alpha} + \upsilon_u R_S^{2\alpha}) R_{S}^{(5+i)\beta}R_{S}^{(5+j)\gamma} \Big]
 \nonumber\\
&&\hspace{1.2cm}
+ \sqrt{2} \kappa_{mij}\kappa_{mkl} \upsilon_{\nu_i^c} R_S^{(5+j)\alpha} R_{S}^{(5+k)\beta}R_{S}^{(5+l)\gamma}
- \frac{1}{3\sqrt{2}} (A_\kappa \kappa)_{ijk} R_S^{(5+i)\alpha} R_{S}^{(5+j)\beta}R_{S}^{(5+k)\gamma}
 \nonumber\\
&&\hspace{1.2cm}
+ \frac{1}{\sqrt{2}}(A_\lambda \lambda)_i R_S^{1\alpha}R_{S}^{2\beta}R_{S}^{(5+i)\gamma}
- \frac{1}{\sqrt{2}}\lambda_i \kappa_{ijk} (\upsilon_u R_S^{1\alpha} + \upsilon_d R_S^{2\alpha}) R_{S}^{(5+j)\beta}R_{S}^{(5+k)\gamma},
\\
&&C_{\alpha \beta \gamma}^{P} = \frac{-e^2}{4\sqrt{2}{s_{_W}^2}{c_{_W}^2}} \Big[ \upsilon_d R_S^{1\alpha}R_{P}^{1\beta}R_{P}^{1\gamma}
+ \upsilon_u R_S^{2\alpha}R_{P}^{2\beta}R_{P}^{2\gamma}
+(\upsilon_d R_S^{1\alpha} + \upsilon_u R_S^{2\alpha}) R_{P}^{(2+i)\beta}R_{P}^{(2+i)\gamma}  \Big]  \nonumber\\
&&\hspace{1.2cm}
+ \frac{1}{\sqrt{2}} \Big[ \lambda_i \lambda_i (\upsilon_d R_S^{1\alpha}R_{P}^{2\beta}R_{P}^{2\gamma}
+ \upsilon_u R_S^{2\alpha}R_{P}^{1\beta}R_{P}^{1\gamma})
- \lambda_i \lambda_j (\upsilon_d R_S^{1\alpha} + \upsilon_u R_S^{2\alpha}) R_{P}^{(5+i)\beta}R_{P}^{(5+j)\gamma} \Big]
 \nonumber\\
&&\hspace{1.2cm}
+ \sqrt{2} \kappa_{mij}\kappa_{mkl} \upsilon_{\nu_i^c} R_S^{(5+l)\alpha} R_{P}^{(5+j)\beta}R_{P}^{(5+k)\gamma}
+ \frac{1}{\sqrt{2}} (A_\kappa \kappa)_{ijk} R_S^{(5+i)\alpha} R_{P}^{(5+j)\beta}R_{P}^{(5+k)\gamma}
 \nonumber\\
&&\hspace{1.2cm}
- \frac{1}{\sqrt{2}}(A_\lambda \lambda)_i \Big[  R_S^{1\alpha}R_{P}^{2\beta}R_{P}^{(5+i)\gamma}
+R_S^{2\alpha}R_{P}^{1\beta}R_{P}^{(5+i)\gamma} + R_S^{(5+i)\alpha}R_{P}^{1\beta}R_{P}^{2\gamma} \Big]
 \nonumber\\
&&\hspace{1.2cm}
+ \frac{1}{\sqrt{2}}\lambda_i \kappa_{ijk} (\upsilon_u R_S^{1\alpha} + \upsilon_d R_S^{2\alpha}) R_{P}^{(5+j)\beta}R_{P}^{(5+k)\gamma}
,
\end{eqnarray}
where the unitary matrices $R_S$, $R_P$ (and $Z_n$, $Z_-$, $Z_+$ below) can be found in Ref.~\cite{ref-zhang1}, and the small terms containing $Y_{\nu_i} \sim \mathcal{O}(10^{-7})$ and $\upsilon_{\nu_i} \sim \mathcal{O}(10^{-4}\,{\rm{GeV}})$ are ignored.

The interaction Lagrangian between CP-even neutral scalars and neutral fermions is formulated as
\begin{eqnarray}
\mathcal{L}_{int} =  S_\alpha \bar{\chi}_\varsigma^0 \Big(C_{L}^{S_\alpha \chi_\eta^0 \bar{\chi}_\varsigma^0}{P_L} + C_{R}^{S_\alpha \chi_\eta^0 \bar{\chi}_\varsigma^0}{P_R}\Big) \chi_\eta^0 ,
\end{eqnarray}
where
\begin{eqnarray}
&&C_{L}^{S_\alpha \chi_\eta^0 \bar{\chi}_\varsigma^0} =  \frac{-e}{2s_{_W}c_{_W}}
\Big(c_{_W}Z_n^{2\eta } - s_{_W}Z_n^{1\eta }\Big)
 \Big( R_S^{1\alpha }Z_n^{3\varsigma} - R_S^{2\alpha }Z_n^{4\varsigma} + R_S^{(2 + i)\alpha} Z_n^{(7 + i)\varsigma}\Big) \nonumber\\
&&\hspace{1.9cm}
- \frac{1}{\sqrt{2}}{Y_{\nu_{ij}}}\Big( R_S^{2\alpha }Z_n^{(7+i)\eta}Z_n^{(4+j)\varsigma} +R_S^{(2+i)\alpha }Z_n^{3\eta}Z_n^{(4+j)\varsigma}  + R_S^{(5+j)\alpha } Z_n^{3\eta}Z_n^{(7+i)\varsigma} \Big) \nonumber\\
&&\hspace{1.9cm}
-\frac{1}{\sqrt{2}}{\lambda_i}\Big( R_S^{1\alpha }Z_n^{(4+i)\eta}Z_n^{4\varsigma} + R_S^{2\alpha }Z_n^{(4+i)\eta}Z_n^{3\varsigma} +  R_S^{(5+i)\alpha }Z_n^{3\eta}Z_n^{4\varsigma}  \Big) \nonumber\\
&&\hspace{1.9cm}
+\frac{1}{\sqrt{2}}{\kappa_{ijk}} R_S^{(5+i)\alpha }Z_n^{(4+j)\eta}Z_n^{(4+k)\varsigma} ,
\\
&&
C_{R}^{S_\alpha \chi_\eta^0 \bar{\chi}_\varsigma^0} = \Big[ C_{L}^{S_\alpha \chi_\varsigma^0 \bar{\chi}_\eta^0} \Big]^ *,
\end{eqnarray}
and
\begin{eqnarray}
P_L=\frac{1}{2}{(1 - {\gamma^5})},\qquad  P_R=\frac{1}{2}{(1 + {\gamma^5})}.
\end{eqnarray}

The interaction Lagrangian of neutral scalars and charged fermions can be written as
\begin{eqnarray}
&&\mathcal{L}_{int} = S_\alpha \bar{\chi}_\zeta (C_L^{{S_\alpha }{\chi _\beta }{{\bar \chi }_\zeta }}{P_L} + C_R^{{S_\alpha }{\chi _\beta }{{\bar \chi }_\zeta }}{P_R}) \chi_\beta + P_\alpha \bar{\chi}_\zeta (C_L^{{P_\alpha }{\chi _\beta }{{\bar \chi }_\zeta }}{P_L}  + C_R^{{P_\alpha }{\chi _\beta }{{\bar \chi }_\zeta }}P_R ) \chi_\beta,
\end{eqnarray}
where the coefficients are
\begin{eqnarray}
&&C_L^{{S_\alpha }{\chi _\beta }{{\bar \chi }_\zeta }} =   \frac{-e}{{{\sqrt{2}s_{_W}}}}\Big[ R_S^{2\alpha }Z_ - ^{1\beta }Z_ + ^{2\zeta } + R_S^{1\alpha }Z_ - ^{2\beta }Z_ + ^{1\zeta } + R_S^{(2 + i)\alpha }Z_ - ^{(2 + i)\beta }Z_ + ^{1\zeta } \Big]  \nonumber\\
&&\qquad\qquad\quad + \,\frac{1}{\sqrt{2}} {Y_{e_{ij}}}\Big[ R_S^{(2 + i)\alpha }Z_ - ^{2\beta }Z_ + ^{(2 + j)\zeta } - R_S^{1\alpha }Z_ - ^{(2 + i)\beta }Z_ + ^{(2 + j)\zeta }  \Big] \nonumber\\
&&\qquad\qquad\quad - \,\frac{1}{\sqrt{2}}{Y_{\nu_{ij}}}R_S^{(5 + j)\alpha }Z_ - ^{(2 + i)\beta }Z_ + ^{2\zeta } - \frac{1}{\sqrt{2}}{\lambda _i}R_S^{(5 + i)\alpha }Z_ - ^{2\beta }Z_ + ^{2\zeta }  ,\\
&&C_L^{{P_\alpha }{\chi _\beta }{{\bar \chi }_\zeta }} = \frac{{ie}}{{{\sqrt{2}s_{_W}}}}\Big[R_P^{2\alpha }Z_ - ^{1\beta }Z_ + ^{2\zeta } + R_P^{1\alpha }Z_ - ^{2\beta }Z_ + ^{1\zeta } + R_P^{(2 + i)\alpha }Z_ - ^{(2 + i)\beta }Z_ + ^{1\zeta }\Big]  \nonumber\\
&&\qquad\qquad\quad  + \, \frac{i}{\sqrt{2}}{Y_{{e_{ij}}}}\Big[ R_P^{(2 + i)\alpha }Z_ - ^{2\beta }Z_ + ^{(2 + j)\zeta } - R_P^{1\alpha }Z_ - ^{(2 + i)\beta }Z_ + ^{(2 + j)\zeta } \Big] \nonumber\\
&&\qquad\qquad\quad  - \, \frac{i}{\sqrt{2}}{Y_{{\nu _{ij}}}}R_P^{(5 + j)\alpha }Z_ - ^{(2 + i)\beta }Z_ + ^{2\zeta } - \frac{i}{\sqrt{2}}{\lambda _i}R_P^{(5 + i)\alpha }Z_ - ^{2\beta }Z_ + ^{2\zeta } , \\
&&C_R^{{S_\alpha }{\chi _\beta }{{\bar \chi }_\zeta }} = \Big[ {C_L^{{S_\alpha }{\chi _\zeta }{{\bar \chi }_\beta }}} \Big]^ * ,\qquad\quad    C_R^{{P_\alpha }{\chi _\beta }{{\bar \chi }_\zeta }} = \Big[ {C_L^{{P_\alpha }{\chi _\zeta }{{\bar \chi }_\beta }}} \Big]^ * .
\end{eqnarray}

The interaction Lagrangian of charged scalars, charged fermions, and neutral fermions can be similarly written by
\begin{eqnarray}
\mathcal{L}_{int} = S_\alpha^- \bar{\chi}_\beta (C_L^{S_\alpha ^ - \chi _\eta ^0 {{\bar \chi }_\beta }}{P_L} + C_R^{S_\alpha ^ - \chi _\eta ^0 {{\bar \chi }_\beta }}{P_R} ) \chi_\eta^0 + S_\alpha^{-\ast} \bar{\chi}_\eta^0 (C_L^{S_\alpha ^{-\ast} {\chi _\beta }\bar \chi _\eta ^0 }{P_L} + C_R^{S_\alpha ^{-\ast} {\chi _\beta }\bar \chi _\eta ^0 }{P_R} ) \chi_\beta ,
\end{eqnarray}
where
\begin{eqnarray}
&&C_L^{S_\alpha^- \chi _\eta^0 {{\bar{\chi}}_\beta }} =   \frac{-e}{{\sqrt{2} {s_W}{c_W}}}R{_{{S^\pm }}^{2\alpha \ast } }Z_+^{2\beta} \Big[ {{c_W}Z_n^{2\eta } + {s_W}Z_n^{1\eta }} \Big]  - \frac{e}{{{s_W}}}R{_{{S^ \pm }}^{2\alpha\ast } }Z_ + ^{1\beta }Z_n^{4\eta }  \nonumber\\
&&\qquad\qquad\quad - \frac{{\sqrt{2} e}}{{{c_W}}}R{_{{S^\pm }}^{(5 + i)\alpha\ast } }Z_ + ^{(2 + i)\beta }Z_n^{1\eta } + {Y_{\nu_{ij}}}R_{{S^ \pm }}^{(2 + i)\alpha }Z_ + ^{2\beta }Z_n^{(4 + j)\eta } \nonumber\\
&&\qquad\qquad\quad  + \, {Y_{e_{ij}}}Z_ + ^{(2 + j)\beta } \Big[ R_{{S^ \pm }}^{1\alpha }Z_n^{(7 + i)\eta } - R_{{S^ \pm }}^{(2 + i)\alpha }Z_n^{3\eta } \Big] - {\lambda _i}R_{{S^ \pm }}^{1\alpha }Z_ + ^{2\beta }Z_n^{(4 + i)\eta },\\
&&C_L^{S_\alpha ^{-\ast} {\chi _\beta }\bar \chi _\eta ^0 } =   \frac{e}{{\sqrt 2 {s_W}{c_W}}}\Big[ R{{_{{S^ \pm }}^{1\alpha\ast }} }Z_ - ^{2\beta } + R{{_{{S^ \pm }}^{(2 + i)\alpha }}^ * }Z_ - ^{(2 + i)\beta }\Big]\Big[ {c_W}Z_n^{2\eta } + {s_W}Z_n^{1\eta }\Big] \nonumber\\
&&\qquad\qquad\quad\;\; - \frac{e}{{{s_W}}}Z_ - ^{1\beta }\Big[ R{{_{{S^ \pm }}^{1\alpha\ast }} }Z_n^{3\eta } + R{{_{{S^ \pm }}^{(2 + i)\alpha\ast }} }Z_n^{(7 + i)\eta }\Big] + {Y_{\nu_{ij}}}R_{{S^ \pm }}^{2\alpha }Z_ - ^{(2 + i)\beta }Z_n^{(4 + j)\eta }\nonumber\\
&&\qquad\qquad\quad\;\; +\: {Y_{{e_{ij}}}}R_{{S^ \pm }}^{(5 + j)\alpha }\Big[ Z_ - ^{2\beta }Z_n^{(7 + i)\eta } - Z_ - ^{(2 + i)\beta }Z_n^{3\eta } \Big] - {\lambda _i}R_{{S^ \pm }}^{2\alpha }Z_ - ^{2\beta }Z_n^{(4 + i)\eta },\\
&&C_R^{S_\alpha ^ - \chi _\eta ^0 {{\bar \chi }_\beta }} = \Big[ {C_L^{S_\alpha ^{-\ast} {\chi _\beta }\bar \chi _\eta ^0 }} \Big]^ * , \qquad  C_R^{S_\alpha ^{-\ast} {\chi _\beta }\bar \chi _\eta ^0 } =\Big[ {C_L^{S_\alpha ^ - \chi _\eta ^0 {{\bar \chi }_\beta }}} \Big]^ * .
\end{eqnarray}

\section{Muon MDM in the $\mu\nu$SSM\label{app-MDM}}

The muon anomalous magnetic dipole moment (MDM) in the $\mu\nu$SSM can be given as the effective Lagrangian
\begin{eqnarray}
\mathcal{L}_{MDM} =\frac{e} {4 m_\mu} a_\mu \bar{l}_\mu \sigma^{\alpha \beta} l_\mu F_{\alpha\beta},
\end{eqnarray}
where $l_\mu$ denotes the muon which is on-shell, $m_\mu$ is the mass of the muon, $\sigma^{\alpha\beta}=\frac{i}{2}[\gamma^\alpha,\gamma^\beta]$, $F_{\alpha\beta}$ represents the electromagnetic field strength and muon MDM $a_\mu=\frac{1}{2}(g-2)_\mu$. Adopting the effective Lagrangian approach, the MDM of the  muon can be written by~\cite{Feng1,Feng2,Feng3}
\begin{eqnarray}
a_\mu = 4 m_{\mu}^2 \Re{(C_2^R + C_2^{L\ast} + C_6^R)},
\end{eqnarray}
where $\Re(\cdots)$ denotes the operation to take the real part of the complex number,
and $C_{2,6}^{L,R}$ represent the Wilson coefficients of the corresponding effective operators $O_{2,6}^{L,R}$
\begin{eqnarray}
&&O_2^{L,R} = \frac{e Q_f}{{(4 \pi)}^2} \overline{(i \mathcal{D}_\alpha l_\mu )}
\gamma^\alpha F\cdot \sigma P_{L,R} l_\mu, \nonumber\\
&&O_6^{L,R} = \frac{e Q_f m_{\mu}}{{(4 \pi)}^2} \overline{l}_\mu F\cdot \sigma P_{L,R} l_\mu.
\end{eqnarray}

The SUSY corrections of the Wilson coefficients in the $\mu\nu$SSM can be
\begin{eqnarray}
C_{2,6}^{L,R}=C_{2,6}^{{L,R}(n)}+C_{2,6}^{{L,R}(c)}.
\end{eqnarray}
The effective coefficients $C_{2,6}^{{L,R}(n)}$ denote the contributions from the neutralinos $\chi_{\eta}^0$ and the charged scalars $S_{\alpha}^-$ loops
\begin{eqnarray}
&&\hspace{-0.75cm}C_2^{R(n)}=\frac{1}{{m_W^2}}C_L^{S_\alpha ^ - \chi _\eta ^0 {{\bar l}_\mu}}
C_R^{S_\alpha ^{-\ast} {l_\mu}\bar \chi _\eta ^0 }\Big[  - {I_3}{\rm{(}}{x_{\chi _\eta ^0 }},{x_{S_\alpha ^ - }}) \nonumber\\
&&\hspace{0.6cm}+ {I_4}({x_{\chi _\eta ^0 }},{x_{S_\alpha ^ - }}) \Big],
\nonumber\\
&&\hspace{-0.75cm}C_6^{R(n)}=\frac{ {m_{\chi _\eta ^0 }}}{{m_W^2}{m_\mu }}
C_R^{S_\alpha ^ - \chi _\eta ^0 {{\bar l}_\mu}}C_R^{S_\alpha ^{-\ast} {l_\mu}\bar \chi _\eta^0 }
\Big[ - 2 {I_1}({x_{\chi _\eta ^ 0 }},{x_{S_\alpha ^ - }}) \nonumber\\
&&\hspace{0.6cm}+ 2 {I_3}({x_{\chi _\eta ^ 0 }},{x_{S_\alpha ^ - }}) \Big],  \nonumber\\
&&\hspace{-0.75cm}C_{2,6}^{L(n)}=C_{2,6}^{R(n)}\mid _{L \leftrightarrow R}.
\end{eqnarray}
Similarly, the contributions $C_{2,6}^{{L,R}(c)}$ coming from the charginos $\chi_{\beta}$ and the neutral scalars $N_{\alpha}$ ($N=S,P$) loops are
\begin{eqnarray}
&&\hspace{-0.75cm}C_2^{R(c)}= \sum\limits_{N=S,P} \frac{1}{{m_W^2}} C_R^{{N_\alpha }{\chi _\beta }{{\bar l }_\mu}}C_L^{{N_\alpha }{l_\mu}{{\bar \chi }_\beta }}\Big[  - {I_1}({x_{{\chi _\beta }}},{x_{{N_\alpha }}}) \nonumber\\
&&\hspace{0.6cm}+ 2{I_3}({x_{{\chi _\beta }}},{x_{{N_\alpha }}})
 - {I_4}({x_{{\chi _\beta }}},{x_{{N_\alpha }}}) \Big],
\nonumber\\
&&\hspace{-0.75cm}C_6^{R(c)}= \sum\limits_{N=S,P} \frac{ {m_{{\chi _{^\beta }}}}}{ {m_W^2}{m_\mu }}
C_R^{{N_\alpha }{\chi _\beta }{{\bar l }_\mu}}C_R^{{N_\alpha }{l _\mu}{{\bar \chi }_\beta }}\Big[ 2 {I_1}({x_{{\chi _\beta }}},{x_{{N_\alpha }}}) \nonumber\\
&&\hspace{0.6cm}- 2 {I_2}({x_{{\chi _\beta }}},{x_{{N_\alpha }}}) - 2 {I_3}({x_{{\chi _\beta }}},{x_{{N_\alpha }}}) \Big] , \nonumber\\
&&\hspace{-0.75cm}C_{2,6}^{L(c)}=C_{2,6}^{R(c)}\mid _{L \leftrightarrow R}.
\end{eqnarray}
Here, the loop functions $I_{i}(\textit{x}_1 , x_2 )$ are given as
\begin{eqnarray}
&&\hspace{-0.75cm}{I_1}(\textit{x}_1 , x_2 ) = \frac{1}{{16{\pi ^2}}}\Big[ \frac{{1 + \ln {x_2}}}{{{x_1} - {x_2}}} - \frac{{{x_1}\ln {x_1}}-{{x_2}\ln {x_2}}}{{{{({x_1} - {x_2})}^2}}} \Big],\\
&&\hspace{-0.75cm}{I_2}(\textit{x}_1 , x_2 ) = \frac{1}{{16{\pi ^2}}}\Big[ - \frac{{1 + \ln {x_1}}}{{{x_1} - {x_2}}} + \frac{{{x_1}\ln {x_1}}-{{x_2}\ln {x_2}}}{{{{({x_1} - {x_2})}^2}}} \Big],\nonumber\\
&&\hspace{-0.75cm}{I_3}(\textit{x}_1 , x_2 ) = \frac{1}{{32{\pi ^2}}}\Big[  \frac{{3 + 2\ln {x_2}}}{{{x_1} - {x_2}}} + \frac{{2{x_2} + 4{x_2}\ln {x_2}}}{{{{({x_1} - {x_2})}^2}}} \nonumber\\
&&\hspace{1.1cm}- \frac{{2x_1^2\ln {x_1}}}{{{{({x_1} - {x_2})}^3}}}  +  \frac{{2x_2^2\ln {x_2}}}{{{{({x_1} - {x_2})}^3}}}\Big], \\
&&\hspace{-0.75cm}{I_4}(\textit{x}_1 , x_2 ) = \frac{1}{{96{\pi ^2}}} \Big[ \frac{{11 + 6\ln {x_2}}}{{{x_1} - {x_2}}} + \frac{{15{x_2} + 18{x_2}\ln {x_2}}}{{{{({x_1} - {x_2})}^2}}} \nonumber\\
&&\hspace{1.1cm}+ \frac{{6x_2^2 + 18x_2^2\ln {x_2}}}{{{{({x_1} - {x_2})}^3}}}  - \frac{{6x_1^3\ln {x_1}}-{6x_2^3\ln {x_2}}}{{{{({x_1} - {x_2})}^4}}}  \Big].\;\;\;\quad
\end{eqnarray}

\end{document}